\documentclass[journal=jacsat,manuscript=article]{achemso}
\setkeys{acs}{keywords = true}
\usepackage{chemformula} 
\usepackage[T1]{fontenc} 

\author{Christian N. Saggau}
\email{c.n.saggau@ifw-dresden.de}
\affiliation{Leibniz Institute for Solid State and Materials Research Dresden (IFW Dresden), 01069 Dresden, Germany}
\altaffiliation{These two authors contributed equally}
  
\author{Sanaz Shokri}
\affiliation{Leibniz Institute for Solid State and Materials Research Dresden (IFW Dresden), 01069 Dresden, Germany}
\alsoaffiliation{Institute of Applied Physics, Technische Universität Dresden, 01062 Dresden, Germany}
\altaffiliation{These two authors contributed equally}

\author{Mickey Martini}
\affiliation{Leibniz Institute for Solid State and Materials Research Dresden (IFW Dresden), 01069 Dresden, Germany}
\alsoaffiliation{Institute of Applied Physics, Technische Universität Dresden, 01062 Dresden, Germany}

\author{Tommaso Confalone}
\affiliation{Leibniz Institute for Solid State and Materials Research Dresden (IFW Dresden), 01069 Dresden, Germany}
\alsoaffiliation{Institute of Applied Physics, Technische Universität Dresden, 01062 Dresden, Germany}

\author{Yejin Lee}
\affiliation{Leibniz Institute for Solid State and Materials Research Dresden (IFW Dresden), 01069 Dresden, Germany}
\alsoaffiliation{Institute of Applied Physics, Technische Universität Dresden, 01062 Dresden, Germany}

\author{Daniel Wolf}
\affiliation{Leibniz Institute for Solid State and Materials Research Dresden (IFW Dresden), 01069 Dresden, Germany}

\author{Genda Gu}
\affiliation{Condensed Matter Physics and Materials Science Department, Brookhaven National Laboratory, Upton, NY 11973, USA}

\author{Valentina Brosco}
\affiliation{Italian National Research Council, Institute for Complex Systems, Rome, Italy}
  
\author{Domenico Montemurro}
\affiliation{Department of Physics, University of Naples Federico II, 80125 Naples, Italy}
  
\author{Valerii\,M.\,Vinokur}
\affiliation{Terra Quantum AG, CH-9000 St.\,Gallen, Switzerland}
\alsoaffiliation{Physics Department, CUNY, City College of City University of New York, 160 Convent Ave, New York, NY 10031, USA}

\author{Kornelius Nielsch}
\affiliation{Leibniz Institute for Solid State and Materials Research Dresden (IFW Dresden), 01069 Dresden, Germany}
\alsoaffiliation{Institute of Applied Physics, Technische Universität Dresden, 01062 Dresden, Germany}
\alsoaffiliation{Institute of Materials Science, Technische Universität Dresden, 01062 Dresden, Germany}
 
\author{Nicola Poccia}
\email{n.poccia@ifw-dresden.de}
\affiliation{Leibniz Institute for Solid State and Materials Research Dresden (IFW Dresden), 01069 Dresden, Germany}

\title[An \textsf{achemso} demo]
  {2D high temperature superconductor integration in contact printed circuit boards}


\begin{document}

\begin{tocentry}

\includegraphics[width=\textwidth]{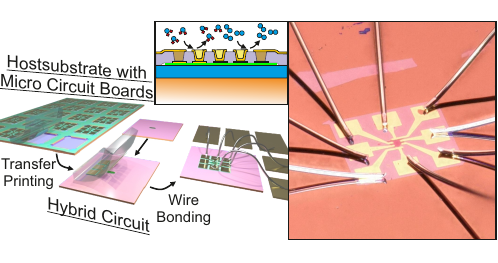}

\end{tocentry}

\begin{abstract}
Inherent properties of superconducting\, Bi$_2$Sr$_2$CaCu$_2$O$_{8+x}$ films, such as the high superconducting transition temperature $T_{\mathrm c}$, efficient Josephson coupling between neighboring CuO layers, and fast quasiparticle relaxation dynamics, make them a promising platform for advances in quantum computing and communication technologies. However, preserving two-dimensional superconductivity during device fabrication is an outstanding experimental challenge because of the fast degradation of the superconducting properties of two-dimensional flakes when they are exposed to moisture, organic solvents, and heat. Here, to realize  superconducting devices utilizing two-dimensional superconducting films, we develop a novel fabrication technique relying on the cryogenic dry transfer of printable circuits embedded into a silicon nitride membrane. This approach separates the circuit fabrication stage requiring chemically reactive substances and ionizing physical processes from the creation of the thin superconducting structures. Apart from providing electrical contacts in a single transfer step, the membrane encapsulates the surface of the crystal shielding it from the environment. The fabricated atomically thin Bi$_2$Sr$_2$CaCu$_2$O$_{8+x}$-based devices show  high superconducting transition temperature $T_{\mathrm c}\simeq 91\,\text{K}$ close to that of the bulk crystal and demonstrate stable superconducting properties.
\end{abstract}

\section{Introduction}
The Bi$_2$Sr$_2$CaCu$_2$O$_{8+x}$ (BSCCO) cuprate having the $d$-wave pairing symmetry\,\cite{tsuei2000pairing} and displaying superconductivity even when being a monolayer ,\,\cite{bib16,bib22} is an ideal candidate for building highly functional quantum devices. Twisted van der Waals (vdW) heterostructures composed of BSCCO layers were predicted\,\cite{can2021high,bib42,liu2023making} to break time-reversal symmetry at certain twist angles, a phenomenon which was demonstrated experimentally.\cite{bib43} As such, BSCCO twisted heterostructures could potentially lead to topological states and Majorana modes\,\cite{mercado2022high} Additionally, twisted interfaces between cuprate crystals have been proposed as an adaptable platform for Josephson junctions with controllable coupling,\,\cite{bib40,bib41,volkov2023current,volkov2023magic,bib25,martini2023twisted} which is a desirable feature for quantum applications.  Thin BSCCO films promise to become an important platform for quantum sensors and detectors, since bolometers fabricated from thin structured BSCCO crystals demonstrated a high level of single photon sensitivity above 25\,K.\,\cite{bib32,bib33} Yet, the main challenge for building these structures is to maintain the superconducting phase of thin BSCCO crystals while integrating them into an electrical circuit, as the oxygen dopants are mobile at temperatures exceeding 200\,K.\,\cite{bib45, bib46} Furthermore, thin BSCCO flakes easily react with water leading to structural changes and modification of the oxygen doping level, which is crucial since altering the doping can drive BSCCO film into an insulating phase.\,\cite{bib14,bib15, bib44, bib56} The preservation of interface superconductivity of the BSCCO crystal has been recently achieved by means of a cryogenic exfoliation technique combined with a solution-free stencil mask approach, enabling the fabrication of electrical devices entirely in an inert atmosphere.\,\cite{bib34,martini2023twisted,bib25} The bottleneck {hindering the realization of functional integrated circuits based on cuprate superconductors is the metallization of atomically thin crystals. The previously mentioned stencil mask approach enables the realization of simple electrical contacts

\begin{figure}
 \includegraphics[width=1\textwidth]{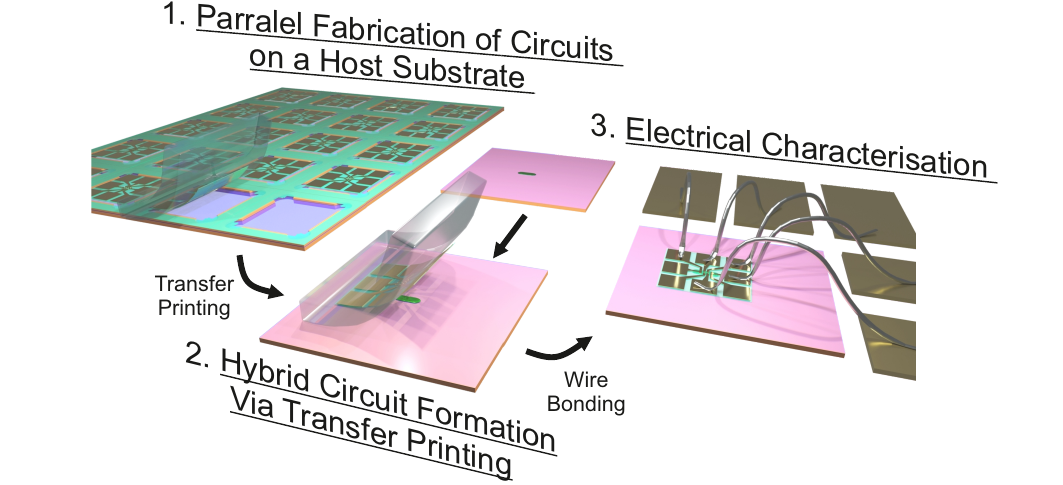}
	\caption{Schematic illustration of the general process flow. 1. Microelectronic circuit boards are fabricated with classical CMOS processes in a highly parallel fashion. 2. Embedded via contacts enable hybrid devices through contact printing of the circuits onto a 2D material. 3. Direct wire bonding facilitates a fast and easy characterization.} 
 \label{fig:Concept}
\end{figure}

\noindent in a single-layer geometry through the thermal evaporation of relatively low melting metals, such as gold, while providing a resolution down to 1 \textmu m.\,\cite{bib22} However,  highly energetic metal clusters and the atoms bombarding the surface of the crystal during the deposition alter its doping level near the contact region due to local heating. To circumvent crystal distortions associated with the metallization process, different cold lamination techniques were developed for other vdW materials. These include the direct transfer of metal films,\,\cite{bib4} contacts attached to the bottom of poly(methyl methacrylate) (PMMA) sheets\,\cite{bib3, bib5}, metal deposition by inkjet printing,\,\cite{bib8} and contacts integrated into the $h$-BN flakes\,\cite{bib7, bib6}. However, all these methods have certain limitations making them unsuitable for integrating complex circuits with materials as sensitive as BSCCO. The first two approaches are limited to simple contact geometries with a minimal feature size around 3\,\textmu m.\cite{bib3, bib4, bib5} Similarly, printed contacts have a comparably low resolution in the range of tens of microns and require an annealing step to obtain an acceptable electrical conductivity.\,\cite{bib9} The BSCCO itself can be damaged during the annealing or printing due to heating or chemical reactions with the binders and solvents of the ink. In contrast, via contacts embedded within $h$-BN flakes can provide a submicron resolution and avoid direct lithographic patterning onto the 2D crystal.\,\cite{bib7, bib6} Nevertheless, the $h$-BN flakes cannot host complex circuits as a consequence of their small size, requiring additional time-consuming processes to obtain a functional device.\,\cite{bib7, bib6} To overcome these limitations, we develop a novel via contact method that relies on the dry cryogenic placement with a polydimethylsiloxane (PDMS) stamp,\,\cite{meitl_2006} to transfer electrical circuits embedded within inorganic SiN\textsubscript{x} nanomembranes. We employ this experimental technique to fabricate and measure a Hall bar device, based on an atomically thin BSCCO crystal.  We show that this technology offers a unique opportunity to introduce cuprate heterostructures in complex circuits, such as microwave resonant superconducting circuits, widely used in the field of optoelectronics\,\cite{day2003broadband,natarajan2012superconducting}, and quantum computing.\,\cite{blais2004cavity}

\section{Experimental}

\subsection{Contact Printed Micro Circuit Boards}\label{Sect:fabrication}

\begin{figure}
  \includegraphics[width=1.0\textwidth]{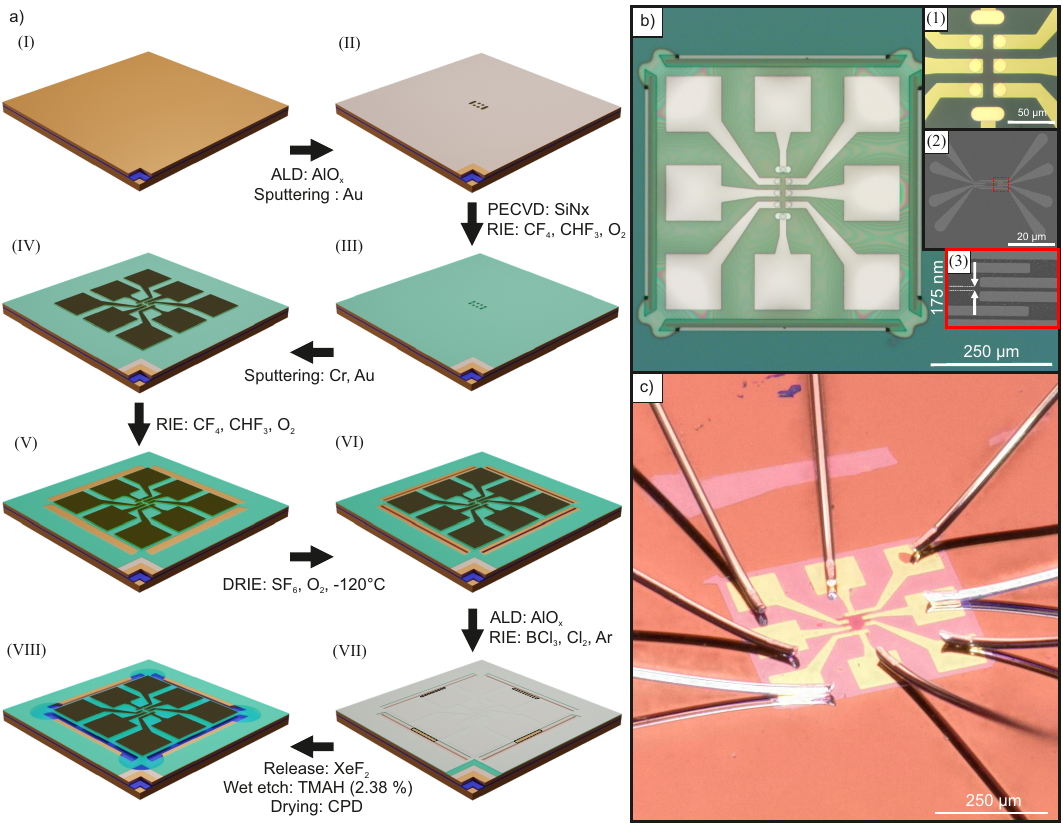}
  \caption{a) Schematic illustration of the fabrication process. (I)\,The SOI substrate; (II)\,The Al\textsubscript{2}O\textsubscript{3} passivated substrate with Au bottom contacts; (III)\,The SiN\textsubscript{x} deposition and via etching; (IV)\, The contact pad definition and via filling; (V)\,The definition of the membrane by RIE and (VI) DRIE; (VII)\,The passivation of the top surface and opening of access windows; (VIII)\,The release of the meambrane and the isotropic removal of the passivation layer. b)\,Microscopic image of a released microcircuit board, the sub-panels (1)-(3), show the bottom side of membranes. The bottom contacts can be distinguished via their difference in color. Sub-panel (1) shows optical lithographically patterned structures, while sub-panel (2) shows an SEM image of the EBL written one. Sub-pannel (3) shows the magnification of the bottom contacts shown in b)(2). c) Microscopy image of a bonded Hall bar device.}
  \label{fig:Fabrication}
\end{figure}

The SiN\textsubscript{x} membrane acts as a transferable microcircuit board (\textmu CB) hosting all the necessary electrical circuitry. The contact formation is completely decoupled from the realization of the electrical contact lines and dielectric barriers that are produced by traditional complementary metal-oxide-semiconductor (CMOS) processes. A 2D thin crystal is electrically contacted by bottom contacts, which are connected through the via contacts with the contacting pads and contact lines at the top surface of the membrane. Upon contact printing of the membrane, as illustrated in Figure\,\ref{fig:Concept}, the device can be directly wire-bonded to a standard chip carrier enabling its electrical characterization. In addition to providing electrical contacts, the membrane acts as a barrier to the environment, facilitating the handling of the integrated 2D material. 

The fabrication of the membranes starts with a silicon-on-insulator (SOI) substrate (Figure\,\ref{fig:Fabrication}a I), where the silicon device layer acts as a sacrificial layer later on, enabling the creation of a freestanding transferable circuit. A 5\,nm thick Al\textsubscript{2}O\textsubscript{3} layer is deposited by the atomic layer deposition (50 cycles ALD) onto the cleaned surface of the substrate. The bottom metal contacts are created, in the following step, on top of the Al\textsubscript{2}O\textsubscript{3} layer, by UV- or electron beam lithography (EBL), sputter deposition, and lift-off (Figure\,\ref{fig:Fabrication}a II). The whole sample is subsequently covered by a 500\,nm thick SiN\textsubscript{x} layer deposited by the plasma enhanced chemical vapor deposition (PECVD). A similar dielectric stack was used previously as a strain engineering platform to form three-dimensional devices, such as capacitors and optical resonators\,\cite{bib10, bib11}. After the deposition of the SiN\textsubscript{x} layer, the via contacts connecting the bottom and the top surface of the membrane are formed. They are fabricated by a two-step process. First, holes are etched by reactive ion etching (RIE) reaching the bottom metal layers (Figure\,\ref{fig:Fabrication}a III). Then the via and electrical conduction lines are created on the top surface using lithography and lift-off (Figure\,\ref{fig:Fabrication}a IV). In the following step, the SiN\textsubscript{x} film is structured to form quadratic individually releasable membranes with a side length of 550\,{\textmu}m hosting tethers at each corner. The tethers are needed to keep the membranes in place once they are released. Additionally, they facilitate their pickup with a polymer stamp (Figure\,\ref{fig:Fabrication}a V). The individual membranes are then isolated from each other through the creation of 5\,{\textmu}m wide trenches reaching the buried oxide (BOX) layer. These trenches are formed in the bare silicon by deep reactive ion etching (DRIE). The whole sample is afterward passivized with a 5\,nm thin Al\textsubscript{2}O\textsubscript{3} layer (50 cycles ALD). The coated trenches suppress the side-wards propagation of the etch front, limiting the release to a restricted area below the \textmu CB's. In the following step, access windows are created in the Al\textsubscript{2}O\textsubscript{3} layer (RIE: BCl\textsubscript{3}, Cl\textsubscript{2}, Ar), enabling XeF\textsubscript{2} gas to access the Si sacrificial layer (Fig \ref{fig:Fabrication}a VII). After the release is finished, the whole chip is immersed for 5 min into a tetramethylammonium hydroxide (TMAH) solution to dissolve the Al\textsubscript{2}O\textsubscript{3} passivation. After that, the chip is transferred to isopropanol and critical point dried (CPD), see Figure \ref{fig:Fabrication}a VIII. After CPD, the chip hosting the \textmu CB is ready to use (More details on the fabrication can be found in the supporting information Figure S2). A freestanding, transferable circuit hosting a Hall bar structure is shown in Figure\,\ref{fig:Fabrication}b. The central bottom side area of a membrane is displayed in Figure\,\ref{fig:Fabrication} b) 1, while it is attached to a PDMS stamp. The bottom contacts, which recline on the surface of the 2D material under investigation, are clearly visible. For the EBL exposed structures, 500\,nm wide contact lines with the sub 200\,nm resolution are obtained, demonstrating a clear advantage in comparison with previous approaches for contact printed contacts, see Figure\,\ref{fig:Fabrication}b) 1-3. Figure \ref{fig:Fabrication}c displays a representative bonded device based on a thin BSCCO crystal. To reveal the sharpness of the interface between the crystal and the electrical contact at atomic resolution, we perform cross-sectional high-annular dark-field scanning transmission electron microscopy (HAADF-STEM) in the region indicated in Figure\,\ref{fig:TEM}a. The TEM image is shown in Figs.\,\ref{fig:TEM}b-c with different magnifications. The gold layer reclines perfectly on the thin crystal displaying a flat edge. Bright spots correspond to the heavy Bi atoms that terminate each layer of the BSCCO crystal. At the top and bottom surfaces of the BSCCO crystal, a sub-unit cell degraded region (dark area) can be observed due to a 30-minute exposure of the flake to the glovebox atmosphere (Ar: 99.999\,\%, O\textsubscript{2} $\leq 0.1 \text{ppm}$, H\textsubscript{2}O $\leq 0.1 \text{ppm}$) between the crystal exfoliation and the stacking of the membrane on top of it. In our system, we are able to preserve interface superconductivity and crystalline order in regions exposed less than one minute to the Ar atmosphere at cryogenic temperatures\,\cite{bib25,martini2023twisted}. A pristine BSCCO/Au interface can thus be created by a quick stacking of the circuit on top of the crystal, which is, however beyond the scope of the present study. 

\begin{figure}
\includegraphics[width=1\textwidth]{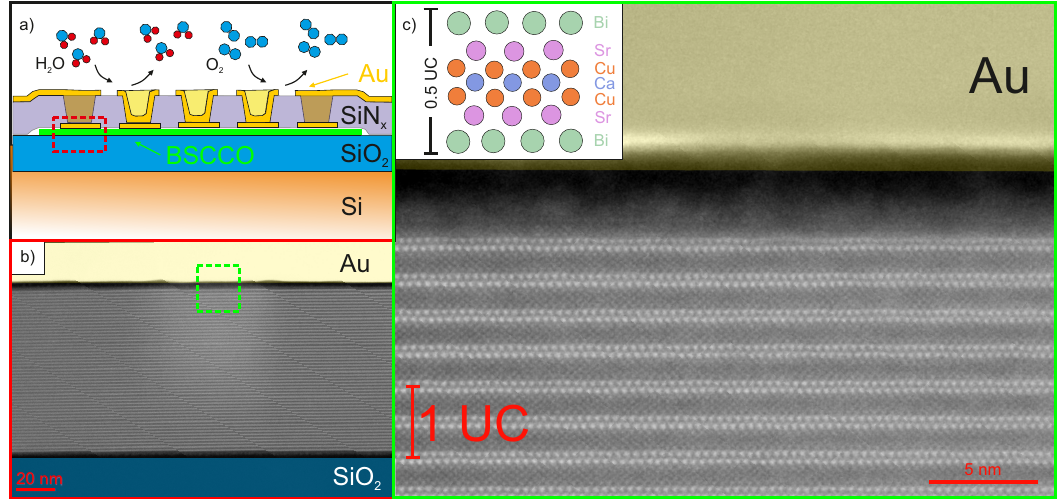}
\caption{a) A sketch of a cut through the central part of the membrane illustrating the via and bottom contacts. The red dashed rectangle indicates the region investigated by cross-sectional high-angle annular dark-field scanning transmission electron microscopy (HAADF-STEM). b) HAADF-STEM image of a BSCCO/Au interface. c)  Zoom into the region within the dashed green rectangle in b). The Bi atoms are most notably resolved as bright spots due to their high atomic mass.} 
\label{fig:TEM}
\end{figure}

\section{Results and discussion}

\subsection{Superconducting Hall Devices}\label{sec2}

To demonstrate the strength of this novel via contact technique, a BSCCO Hall device is fabricated from an atomically thin crystal. The schematic illustration of the fabrication process, along with experimental details, is reported in Figure\,S2 in the Supplementary Information (SI). In the first place, the chip hosting the \textmu CBs is placed on a liquid nitrogen-cooled stage kept at -30\,\textdegree C A PDMS stamp is then brought into contact with a \textmu CB, utilizing a micro-manipulator, and let it thermalize to enhance its stickiness. By quickly detaching the stamp from the chip, we pick up the target membrane hosting the circuit. Next, the BSCCO flakes are exfoliated at room temperature via the scotch tape method on a SiO\textsubscript{2}/Si substrate, previously treated with oxygen plasma to enhance the vdW forces between BSCCO and SiO\textsubscript{2}, and baked overnight (180\,\textdegree C)  to get rid of water molecules. By optical contrast, we identify a very thin flake ($\simeq \text{13.5} \text{nm}$), as shown in Figure\,\ref{fig:Hall_DEvice2}a, and align the membrane before the contact printing. Finally, the sample is slowly heated up to room temperature, followed by the slow removal of the PDMS stamp. One of the key advantages of our technique is that the electrical contacts are established at low temperatures within a few minutes after 

\begin{figure}
	\includegraphics[width=1.0\textwidth]{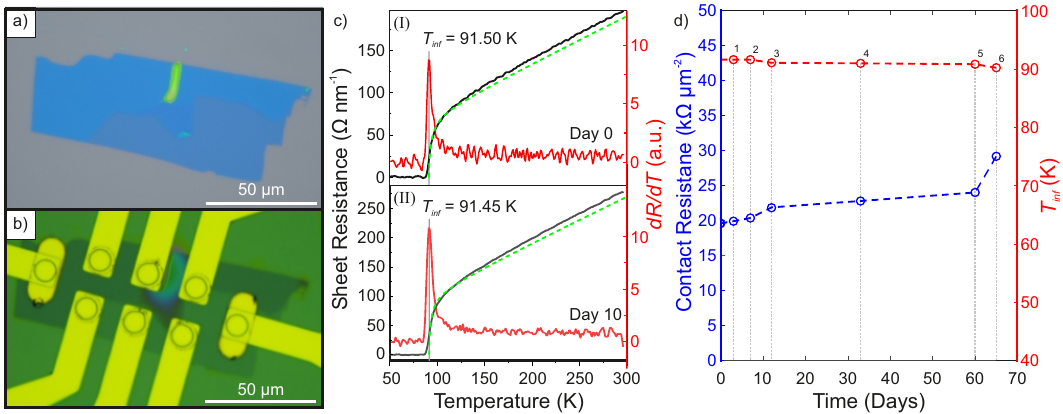}
	\caption{a) Microscopy image of the BSCCO crystal with a thickness of $\simeq$13.5\,nm (4.5 unit cells). b) Closeup of the the device consisting of the flake contacted by a \textmu CB. c) Sheet resistance (left axis) of the crystal across the superconducting transition measured right after the fabrication of the device (I) and after ten days (II). The black solid and green dashed lines correspond to the experimental data and the fit, respectively. The red curves (right axis) represent the derivative of the experimental data displaying the inflection point $T_{inf}$. The fitting parameters are $\tau_\varphi=0.38\,\text{ns}$, $\tau=0.1\,\text{ps}$, $T_{\mathrm c}=91.28\,\text{K}$ for case (I); $\tau_\varphi=0.36\,\text{ns}$, $\tau=0.1\,\text{ps}$, $T_{\mathrm c}=91.39\,\text{K}$ for case\,(II). d) Mean areal contact resistance at room temperature (left axis) and the inflection point $T_{inf}$ (right axis) of a 40\,nm-thick BSCCO flake measured over time. Between each measurement the sample is stored in an Ar atmosphere and kept at -40$^{\circ}$\,C between \#1 and \#5 and at room temperature between \#5 and \#6.
 \label{fig:Hall_DEvice2}}
\end{figure}

\noindent the crystals cleavage before the top layers of the cuprate crystal strongly degrade. Figure\,\ref{fig:Hall_DEvice2}b displays the close-up of the contact area of the device right after printing the \textmu CB onto the BSCCO flake. Our protocol allows for extremely clean electrical contacts to BSCCO with an areal resistance $<$ 35\,k$\Omega$ \textmu m\textsuperscript{-2}. The observed contact resistances are similar to the one obtained by evaporation, while for transfer printed contacts a lower parameter spread is observed (Figure S5). Additionally, evaporated, low ohmic contacts sometimes show indications of degradation, which we do not observe for transfer printed one. For example, some samples with evaporated contacts show a reminiscent low resistance below $T_c$ an increase of the resistance upon reaching a resistance below the noise floor, if the temperature is further decreased. Occasionally no superconducting transition is observed at all. The high electronic quality of our fabricated device is also revealed in Figure\,\ref{fig:Hall_DEvice2}c, which reports the 4-point sheet resistance for the Hall device measured right after its fabrication (I) and after storing the sample 10 days in an Ar filled glovebox at -40\,\textdegree C  (II). The two data sets show nearly identical inflection points of 91.45\,K and 91.50\,K, respectively. The small changes in the  $T_c$ and the sheet resistance can be explained by the loss of oxygen dopants, as even a storage at - 40\,\textdegree C cannot totally prevent a degradation of the sample over time. To totally stop the loss of dopants, even lower temperatures are required, as the mobile oxygen is frozen out at temperatures below 200\,K \,\cite{bib45, bib46}. To arrive at a consistent theoretical description, the behavior of the system across the metal-insulator transition (SIT) is analyzed employing the theory of two-dimensional superconducting fluctuations\,\cite{bib22, bib36}. The scattering time $\tau$ is taken as approximately 0.1\,ps, while the inelastic scattering time $\tau_\varphi$ and the critical temperature $T_c$ are extracted from the fits. For the two sets of data (see Tab.\,\ref{tab1}), we get similar results, with a slightly larger $T_c$ and $\tau_\varphi$ observed for the data shown in Figure\,\ref{fig:Hall_DEvice2} c I, the sample being measured immediately after its fabrication. In both cases, the critical temperature estimated from the fits is close to the inflection point of the experimental data, and it is identical to the value reported for the optimally doped bulk crystal\,\cite{bib13}, demonstrating a device with pristine and stable superconducting properties. This is confirmed by monitoring the values of the contact resistance and the critical superconducting temperature for a 40\,nm thick crystal over several weeks (Figure\,\ref{fig:Hall_DEvice2}d). All these findings suggest that our methodology based on the transfer of \textmu CB preserves the superconducting properties of the BSCCO crystal and that the membrane acts as an encapsulation layer blocking detrimental effects of disorder over a long time. The robustness of the membrane is demonstrated by the numerous thermal cycles performed for our experiments. The Hall resistance and the magnetoresistance of the thin device (shown in Figure\,\ref{fig:Hall_DEvice2}a) and the corresponding theoretical analysis are presented in the SI.

\begin{table}
  \caption{Fitting parameters for superconducting transition}
  \label{tab1}
  \begin{tabular}{@{}llll@{}}
    \hline
Set: & $T_c$  & $\tau_\varphi$  & $\tau$ \\
\hline
Fig.\ref{fig:Hall_DEvice2} (I) $~$  & 91.28 K $~$ & 0.38 ns $~$  & 0.1 ps \\
Fig.\ref{fig:Hall_DEvice2} (II) $~$  & 91.39 K $~$ & 0.36 ns $~$  & 0.1 ps \\
    \hline
  \end{tabular}
\end{table}

\section{\label{sec:conclusion}Conclusion}

In summary, we demonstrate a novel route to integrate sensitive and thin BSCCO crystals into complex nanodevices. This integration is achieved by the cryogenic dry transfer of microcircuits embedded in an inorganic dielectric nanomembrane, where the via contacts connect the bottom electrical contacts to the 2D crystal with contact lines formed on the top surface of the membrane. Additionally, the membrane provides a physical barrier to the environment, which enables the handling of sensitive devices in ambient conditions. Employing this fabrication technique, we realize an atomically thin BSCCO device with high-quality electrical contacts (areal resistance <\,35\,k$\Omega$ \textmu m\textsuperscript{-2}), exhibiting pristine and stable superconducting properties for a very long time ($T_C \simeq 91.5$\,K). Our technology, with the capability to realize high-resolution contact lines on multiple levels and through the use of inorganic dielectrics, provides an effective path toward the implementation of electric gates necessary for individually tuneable devices, a core requirement for complex integrated circuits.

\section{\label{sec:methods}Methods}
\noindent \textbf{Micro Circuit Boards:}
The SOI wafers are purchased from \textit{Ultrasil} and diced into 10 X 10\,mm substrates before use. The substrates are cleaned for 10 min under mild ultrasonication in a bath of 1., hot dimethyl sulfoxide (DMSO) (120\,\textdegree C); 2., acetone; 3., isopropanol. The samples are then blown dry under nitrogen flow, and cleaned in an O\textsubscript{2} plasma  for 30 min (Diener Pico). 
\newline Below, we provide the details of each process section of the fabrication:\\

\noindent  \textbf{ALD:} Al\textsubscript{2}O\textsubscript{3} is deposited by the ALD \textit{Arradiance GEMstar} at 280\textdegree C with H\textsubscript{2}O and trimethylaluminum as precursors.

\noindent  \textbf{PECVD:} SiN\textsubscript{x} is deposited by PECVD (Sentech Si500D) at 280\textdegree C with SiH\textsubscript{4} (5\% in He) (50\,sccm) and N\textsubscript{2} (80\,sccm) as precursors. The pressure is $p = 0.019$\,mbar, and the power of the inductively coupled plasma (ICP) $P = 200$\,W.\\

\noindent  \textbf{RIE:} SiN\textsubscript{x} is etched using \textit{Oxford instruments - Plasmalab 80} system. Bias voltage $U = 155$\,V, ICP-Power: $P = 50$\,W, $T = 30$\,\textdegree C, the precursors are CF\textsubscript{4} with a flow of 10\,sccm, CHF\textsubscript{3} at 20\,sccm, O\textsubscript{2} at 4\,sccm.\\

\noindent  \textbf{RIE:} Al\textsubscript{2}O\textsubscript{3} is etched under the same conditions, but with different precursors (BCl\textsubscript{3} at 20\,sccm, Cl\textsubscript{2})\\

\noindent \textbf{RIE:} Si is etched using SF\textsubscript{6} at 80\,sccm and O\textsubscript{2} at 8\,sccm as precursors (ICP-Power: $P = 450$\,W, bias voltage: $U= -30$\,V, $T = 120$\,\textdegree C.\\

\noindent \textbf{Lithography:} Resist AZ5214E is spun onto samples at 4500\,rpm and pre-baked for 180\,s on a hotplate at 90\,\textdegree C. The samples are then exposed to a maskless aligner (\textit{Heidelberg MLA 100}). For all etching steps, the resist is used as a positive (dose $= 90$ mJ\,cm\textsuperscript{-1}). The samples are finally developed for 60\,s with AZ 726 MIF. For the lift-off a negative slope is used (dose $= 30$\,mJ\,cm\textsuperscript{-2}, post-baking for 450\,s at 120\,\textdegree C and flood-exposure for 45\,s). For the definition of the bottom contacts, the development process lasts 45\,s (with AR 300-35), while for top contacts it lasts 16\,s (with AZ 726 MIF).\\

\noindent \textbf{Metallization}: contacts are sputtered using the \textit{Torr CRC600 Series} tool.  Gold bottom contacts are deposited with a rate of \r{A}s\textsuperscript{-1} for a total thickness of 80\,nm. The top contacts are formed by depositing 5\,nm of Cr with a rate of 1 \r{A}s\textsuperscript{-1} and 80\,nm of Au at 1 \r{A}s\textsuperscript{-1}. Lift-off is performed in DMSO overnight.\\

\noindent \textbf{Release}: the samples are exposed in a gas phase etcher (\textit{SPTS, Xactix}) to multiple cycles (60\,s exposure) of XeF\textsubscript{2} vapor (4\,mbar) until the membranes are completely freestanding. The progress of the release is monitored with an optical microscope. The released membranes are immersed into an alkali developer (AZ 726 MIF) for 5\,min to remove to bottom Al\textsubscript{2}O\textsubscript{3} layer, transferred for 10\,min into H\textsubscript{2}O, before being transferred into isopropanol overnight. The prepared sample is then CPD dried and baked overnight on a hotplate at 125\,\textdegree C to remove residual water.\\

\noindent \textbf{Cleaving of BSCCO:} Floating-zone grown $\text{Bi}_{2}\text{Sr}_{2}\text{CaCu}_{2}\text{O}_{8+d}$ crystals cut orthogonal to the $c$-axis are mechanically exfoliated inside a high purity argon filled glove-box (MBraun $ \text{H\textsubscript{2}O} <0.1\,\text{ppm}$, $ \text{O\textsubscript{2}}< 0.1\,\text{ppm}$), onto Si/SiO\textsubscript{2} substrates via the Scotch tape method (3M, Scotch magic tape). Rectangular flakes with a desired size ($L \geq 100\,\text{\textmu m}, w \geq 40 \text{\textmu m}$  and thickness  $h \leq 40~ \text{nm}$), are selected under an optical microscope.\\

\noindent \textbf{Electrical characterisation: }
After transferring a \textmu CB onto a BSCCO thin flake, the sample is glued onto a sample holder with silver glue and bonded under ambient conditions (\textit{FS Bondtec 58 series}). The sample is loaded into a cryostat (\textit{Quantum Design PPMS}) and the electronic characteristics are recorded with a lock-in amplifier (\textit{Stanford Research Systems SR830}). The Hall and magnetoresistance data shown in the SI are symmetrized to exclude geometric effects. The electrical characterization of Figure\,\ref{fig:Hall_DEvice2} d) is performed in a cryostat (\textit{Quantum Design Versalab}). The circuit used to evaluate the contact resistance is shown in Figure S4 and the spread of the contact resistance for contact printed and evaporated contacts is shown in Figure S5. The injected current is $I = 1$\,\textmu A for 1, $I = 5$\,\textmu A for 2 \& 4-6 . For 3 the injected current is $I = 10$\,\textmu A. \\

\begin{acknowledgement}

The authors thank the Deutsche Forschungsgemeinschaft for financial support (DFG 452128813, DFG 512734967, DFG 492704387, DFG 460444718). The work of V.M.V. and C.N.S. was supported by Terra Quantum AG and partly by the NSF 2105048 (V.M.V). The authors are grateful to Heiko Reith and Ronny Engelhard for providing access to cleanroom infrastructure and Nicolas Perez Rodriguez for providing access to cryogenic measurement facilities. The authors thank Henk-Willem Veltkamp for technical support at the MESA+ facility. The authors thank Thomas Wiek for the preparation of the TEM lamella. The work at BNL was supported by the US Department of Energy, Office of Basic Energy Sciences, contract no. DOE-sc0012704. The authors are also grateful to S. Y. Frank Zhao, Philip Kim, Hans Hilgenkamp, and Francesco Tafuri for illuminating and fruitful discussions.\\

\noindent \textbf{Competing Interests}: The authors declare that they
have no competing interests.\\

\noindent \textbf{Author contributions:} N.P. supervised the experiment;
C.N.S. and N.P. conceived the micro-printed contacts
and their cryogenic transfer methodology; C.N.S., S.S., Y.L.,
performed the experiments and analyzed the data with
the contribution of V.B., T.C., and V.M.V.;
D. W. performed the HAADF-STEM analysis; The
theoretical analysis of the Hall effect was done by V.B.
and V.M.V.; The cuprate crystals were provided by G.G.
The fabrication procedure and the results were discussed
by C.N.S., S.S., D.M., V.M.V., K.N., and
N.P. The manuscript was written by C.N.S., M.M.,
V.B., V.M.V., K.N., and N.P.; all authors discussed the
manuscript.

\end{acknowledgement}

\begin{suppinfo}

The following files are available free of charge.
\begin{itemize}
 \item Supporting Information: 2D high temperature superconductor integration in contact printed circuit boards: The supporting information includes a more detailed description of the experimental methodology and fabrication methods. Furthermore a detailed analysis of the Hall data is provided.
\end{itemize}

\end{suppinfo}

\bibliography{achemso-demo}

\end{document}